\documentclass[aps,preprint,nofootinbib]{revtex4}
\usepackage{graphicx}
\usepackage{color}
\usepackage{epsfig}

\begin{document}
\title{Growing length and time scales in glass forming liquids}

\author{Smarajit Karmakar$^{1}$}
\email{smarajit@physics.iisc.ernet.in}
\author{Chandan Dasgupta$^{1,2}$}
\email{cdgupta@physics.iisc.ernet.in}
\author{Srikanth Sastry$^{2}$}
\email{sastry@jncasr.ac.in}

\affiliation{$^1$ Centre for Condensed Matter Theory, Department of Physics, 
Indian Institute of Science, Bangalore, 560012, India}
\affiliation{$^2$ Jawaharlal Nehru Centre for Advanced Scientific Research, Bangalore
560064, India.}

\maketitle 

{\bf The nature of the glass transition, whereby a variety of liquids
transform into amorphous solids at low temperatures, is a subject of
intense research despite decades of investigation. Explaining the
enormous increase in viscosity and relaxation times of a liquid upon
supercooling is essential for an understanding of the glass
transition.  Although the notion of a growing length scale of
``cooperatively rearranging regions'' (CRR), embodied in the
celebrated Adam-Gibbs relation, has been invoked for a long time to
explain dynamical slow down, the role of length scales relevant to
glassy dynamics is not well established.  Recent studies of spatial
heterogeneity in dynamics provide fresh impetus in this direction. A
powerful approach to extract length scales in critical phenomena is
finite size scaling, wherein a system is studied for a range of sizes
traversing the length scales of interest. We perform finite size
scaling for the first time for a realistic glass former, using
computer simulations, to evaluate a length scale associated with
dynamical heterogeneity which grows as temperature decreases.
However, relaxation times which also grow with decreasing temperature,
do not show the same kind of scaling behavior with system size as the
dynamical heterogeneity, indicating that relaxation times are not
solely determined by the length scale of dynamical heterogeneity.  We
show that relaxation times are instead determined, for all studied
system sizes and temperatures, by configurational entropy, in
accordance with the Adam-Gibbs relation, but in disagreement with
theoretical expectations based on spin-glass models that
configurational entropy is not relevant at temperatures substantially 
above the critical temperature 
of mode coupling theory. The temperature dependence of the
heterogeneity lengthscale shows significant deviations from
theoretical expectations, and the length scale one may extract from
the system size dependence of the configurational entropy has much
weaker temperature dependence compared to the heterogeneity length
scale at all studied temperatures. Our results provide new insights
into the dynamics of glass-forming liquids and pose serious challenges
to existing theoretical descriptions.}

Most approaches to understanding the glass transition and slow dynamics
in glass formers  
\cite{adam-gibbs,jack,debstil,Gotze,inhmct,rfotpra,rfotrev,mezard,kinetic} 
are based on the intuitive picture that the movement of their
constituent particles (atoms, molecules, polymers) requires
progressively more {\it cooperative} rearrangement of groups of
particles as temperature decreases (or density increases). Structural
relaxation becomes slow because the concerted motion of many
particles is infrequent. Intuitively, the size of such ``cooperatively
rearranging regions'' (CRR) is expected to increase with decreasing
temperature. Thus, the above picture
naturally involves the notion of a growing length scale, albeit implicitly in
most descriptions. The notion of such a length scale, related to the 
configurational entropy $S_c$ (see methods), forms the basis of 
rationalizing \cite{adam-gibbs,rfotpra,rfotrev} the 
celebrated Adam-Gibbs (AG) relation \cite{adam-gibbs} 
between the relaxation time and $S_c$.  

More recently, a number of theoretical approaches have explored the
relevance of a growing length scale to dynamical slow down
\cite{rfotrev,inhmct,kinetic}. A specific motivation for some of these 
approaches arises from the study of heterogeneous dynamics in glass
formers\cite{edigerhet,onuki,harrowell,donati}. In particular,
computer simulation studies\cite{onuki,harrowell,donati} focused
attention on spatially correlated groups of particles which exhibit
 enhanced mobility, and whose spatial extent grows upon
decreasing temperature. The spatial correlations of local relaxation permits
identification of a dynamical (time dependent) length scale, $\xi$,
through analysis of a {\it four point} correlation function first
introduced by Dasgupta {\it et al} \cite{chandan} (see methods), and the
associated dynamical susceptibility $\chi_4$\cite{silvio,sharon}. These
quantities have been studied recently {\it via} inhomogeneous
mode coupling theory (IMCT) \cite{inhmct} and estimated from simulation and
experimental
data\cite{inhmct,berthier_fss,berthier_wg,berthier_science,berthier_physreve,zamponi}.

The method of finite size scaling, used extensively in numerical studies of
critical phenomena \cite{fss}, is uniquely suited for investigations of the
presence of a dominant length scale. This method involves a study of the
dependence of the properties of a finite system on its size.
We study a binary mixture of particles interacting via the
Lennard-Jones potential\cite{kob}, originally proposed as a model for
$Ni_{80} P_{20}$, and widely studied as a model glass former. We perform
constant temperature molecular dynamics simulations at a constant
volume (see methods and \cite{sastryPRL} for details), for seven
temperatures, and up to a dozen different system sizes for each temperature. 
For each case, we calculate the dynamic susceptibility $\chi_4(t)$
as the second moment of the distribution of 
a correlation function $Q(t)$ which measures the overlap of
the configuration of particles at a given time with the configuration
after a time $t$ (see
methods). From previous work, it is now well-established that
$\chi_4(t)$ has non-monotonic time dependence, and peaks at a time
$\tau_4$ that is proportional to the structural relaxation time
$\tau$. Such behavior is shown in the inset of Figure 1(a). In Figure
1(a), we show the peak values $\chi_4^p \equiv 
\chi_4(\tau_4)$ {\it vs} system size (number of particles)
$N$ for a range of temperatures. At each temperature, $\chi_4^p$
is an increasing function of $N$, saturating at large $N$.
The saturation occurs at a larger value of $N$ at lower temperatures.
This is precisely the finite-size scaling behaviour expected for a quantity whose
growth with decreasing temperature is governed by a  dominant
correlation length that increases with decreasing temperature. 

We have estimated the correlation length $\xi$ from finite size
scaling of $\chi_4^p(T,N)$, which also involves estimating the value
of $\chi_4^p$ as $N \rightarrow \infty$. As the latter estimation is a
potential source of error in estimating $\xi$, we employ the Binder
cumulant of the distribution of $Q(\tau_4)$ to estimate $\xi$.  The
Binder cumulant \cite{binder}, defined (see methods) in terms of the
fourth and second moments of the distribution, vanishes for a Gaussian
distribution, whereas it acquires negative values which are larger for
distributions that deviate from a Gaussian more strongly. The Binder
cumulant has been used extensively in finite size scaling analysis in
the context of critical phenomena, owing to its very useful property
that in systems with a dominant correlation length $\xi$, it is a
scaling function only of $L/\xi$ (or equivalently, of $N/\xi^3$),
where $L$ is the linear dimension of the system. The distributions
themselves are shown in the inset of Figure 1(b), for two different
system sizes for temperature $T = 0.47$. We see that the distribution
is unimodal for the large system size of $N = 1600$ whereas it is
strongly bimodal for the small system size of $N = 150$. The same
trend is observed as temperature is decreased for a fixed size of the
system. The data collapse of the Binder cumulant, from which we
extract the correlation length $\xi(T)$, is shown in Figure 1(b). The
collapse observed is excellent, confirming that the growth of
$\chi_4^p$ with decreasing $T$ is governed by a growing dynamical
correlation length. The values of $\xi$ obtained from this scaling
analysis are consistent with less accurate estimates obtained from a
similar analysis of the $N$-dependence of $\chi_4^p(T,N)$, as well as
from the wave vector dependence of the four point dynamic structure
factor $S_4(q,\tau_4)$ (see, {\it e. g.} \cite{inhmct}). Estimated $\chi_4^p$ as $N \rightarrow
\infty$ compare very well with the $q
\rightarrow 0$ limit of $S_4(q,\tau_4)$, up to a
proportionality constant, implying that concerns about the dependence
of $\chi_4^p$ values on the details of the simulations do not apply in
our case \cite{berthierjcp}.

The value of $\xi$ (up to a temperature-independent multiplicative
factor that can not be determined from scaling, but has been fixed to
match results obtained from $S_4(q,\tau_4)$) increases from $2.11$ to
$6.23$ as $T$ decreases from $T = 0.70$ to $T = 0.45$.  We find that
both $\xi$ and the asymptotic, $N \to \infty$ value of $\chi_4^p$
deviate from power law behavior as the critical temperature $T_c$ of
mode coupling theory ($T_c \simeq 0.435$ in our system) is approached
(consistently with previous observations). However, the power-law
relationship between $\chi_4^p$ and $\xi$ predicted in IMCT is
approximately satisfied by our data. Since the range of the measured
values of $\xi$ is small, it is not possible to obtain accurate
estimates of the exponents of these power laws.

Next we consider the dependence of the relaxation time $\tau$ on $T$
and $N$.  For each case, we calculate the relaxation time from the
decay of $<Q(t)>$.  The results for $\tau$ are displayed in Figure 2,
which shows that $\tau$ increases as the temperature decreases, as
expected. However, the observed {\it increase} in $\tau$ with
decreasing $N$ for small values of $N$ at fixed $T$ is {\it not}
consistent with standard dynamical scaling for a system with a
dominant correlation length ({\it e.g.}  near a critical point):
dynamical finite-size scaling would predict a decrease in $\tau$ as
the linear dimension $L$ of the system is decreased below the
correlation length $\xi$. Similar finite size effects on relaxation
times have been observed in previous simulations of realistic glass
formers ({\it e. g.} \cite{yamamoto}) but have not been analyzed in
detail. The observed $N$-dependence of $\tau$ is opposite to that
found in finite-size scaling studies of mean-field spin-glass models
~\cite{binder_fss,rcr} whose analytically tractable behaviour forms
the starting point of most recent theories (including IMCT) of the
dynamics of glassy liquids \cite{brumer}.

The inset of Figure 3 shows the large-$N$ value of $\tau$ plotted as a
function of (a) the correlation length $\xi$ on a double-log scale, 
and (b) ${\xi \over k_B T}$ on a semi-log scale.  The
power-law relation between these two quantities predicted in IMCT
\cite{inhmct} is found above $T = 0.5$; deviation from a power law is
found at lower temperatures. The semi-log plot indicates that an
exponential form $\tau \sim \exp\left( k (\xi/k_B T)^\zeta\right)$,
with $\zeta = 0.7$, describes the data well in the entire
range. Though such a dependence is expected according to the random
first order theory (RFOT)\cite{rfotrev}, the exponent value we observe
cannot be easily rationalized within that framework. We comment
further on the significance of the exponent value later. Figure 3
shows relaxation times $\tau(T,N)$ for different $N$ values scaled to
the asymptotic $N \rightarrow \infty$ value $\tau(T)$, plotted against
values of $\chi_4^p(T,N)$ scaled to the asymptotic $N \rightarrow
\infty$ value $\chi_4^p(T)$. If the system size dependence of
$\tau$ and $\chi_4^p$ are governed by the same length scale,
 one must expect a universal dependence of the scaled
relaxation times on the scaled $\chi_4^p$ values. From the data shown
in Figure 3, it is clear that there is no systematic relation between
the scaled $\tau$ and $\chi_4^p$ that describes their variation both with $T$ and
with $N$.  These results indicate that the growth of $\tau$ with
decreasing $T$ is {\it not} governed solely by the correlation length
$\xi$ that describes the growth of $\chi_4$.

Motivated by the AG relation \cite{adam-gibbs}, $\tau \propto 
\exp\left({A \over T S_c}\right)$, where $A$ is a constant, we
next consider the dependence of $\tau$ on the configurational entropy
$S_c$ whose evaluation is described elsewhere \cite{sastryPRL}.  As
shown in Figure 4 where $\log(\tau)$ is plotted {\it vs.}
$\frac{1}{TS_c}$ for all temperatures and system sizes studied, we
find a remarkable agreement with the AG relation, not only {\it vs.}
$T$ but also for all system sizes. To our knowledge, such a
demonstration of the validity of the AG relation for finite or
confined systems has not been made earlier. Thus, the $N$-dependence of
$\tau$, which can not be understood from dynamical finite-size
scaling, can be explained in terms of the $N$-dependence of $S_c$,
suggesting that the growth of $\tau$ with decreasing temperature is
more intimately related to the change of $S_c$, than to the increase
of the correlation length $\xi$ and susceptibility $\chi_4$,
associated with dynamical heterogeneity. As $S_c$ at a given
temperature varies with system size $N$, it is tempting to inquire if
the $N$-dependence of $S_c$ is associated with a length scale. We
extract such a length scale from data collapse of $S_c(T,N)$, scaled
to its value as $N
\rightarrow \infty$, shown in the top inset of Figure 4.  We obtain
reasonable data collapse, but the extracted length scales turn out to
have substantially weaker $T$-dependence compared to $\xi$. The
comparison of the length scale obtained from the configurational
entropy and that obtained from dynamical heterogeneity is shown in the
bottom inset of Figure 4.

A central role for the configurational entropy, along with an analysis
of a length scale relevant to structural relaxation, are the content
of the random first order theory, developed by Wolynes and co-workers
\cite{rfotrev}. According to RFOT, the length scale of dynamical 
heterogeneity is the ``mosaic length'' that represents the critical
size for entropy driven nucleation of a new structure in a
liquid. Mean-field arguments based on known properties of
infinite-range models suggest that the RFOT mechanism is operative for
temperatures lower than $T_{MCT}$. In this regime, the dynamics of the
system is activated, with the relaxation time expected to vary as $\tau =
\tau_0 \exp\left[B \left(\Delta F \over k_B T\right)^\psi\right]$, where 
$\Delta F$ is the free energy barrier to structural rearrangements,
and $\psi$ is an unknown exponent. The free energy barrier in turn
depends on the mosaic length as ${\Delta F \over k_B T} \sim 
\xi^\theta$, where $\theta$ describes the dependence of the surface
energy on the size of a region undergoing structural change. The
observed validity of the AG relation, and the dependence of the
relaxation time on the length scale $\xi$, $\tau \sim \exp\left( k
(\xi/k_B T)^\zeta\right)$, with $\zeta = 0.7$, can be rationalized
within RFOT if the exponent $\theta$ is assumed to be close to 2.3,
and the exponent $\psi$ is close to 0.3. However, this interpretation
has the drawback that the exponent $\theta$ does not satisfy the
physical bound, $\theta \leq 2$, in three dimensions, and there is no
evident explanation for the value of $\psi$. We note that similar
conclusions were reached in a recent analysis \cite{zamponi} of
experimental data near the laboratory glass transition, on a large
class of glass-forming materials. Thus we find puzzling values for 
the exponents relevant to the applicability of RFOT, which are in 
need of explanation, and data in \cite{zamponi} indicate that such 
a result may apply for a wide range of temperatures, all the way to 
the experimental glass transition. 

RFOT focuses on behavior near the glass transition, and in the
limiting case of the spin glass models where theoretical perditions
are available, configurational entropy plays no role in the behavior
of the system above the mode coupling temperature. However, there have
indeed been attempts to extend the RFOT analysis to temperatures above
the mode coupling temperature\cite{stevenson,silvio_kac} and to
estimate a mosaic length scale at such temperatures, and we thus
compare our results with predictions arising from these analyses.
Stevenson {\it et al}\cite{stevenson} have considered the change in
morphology of rearranging regions above the mode coupling temperature,
and correspondingly the dependence of relaxation times on
configurational entropy. The predicted dependence of relaxation times
on configurational entropy differs from the Adam-Gibbs form, whereas
our results strikingly confirm the Adam-Gibbs form. Franz and
Montanari \cite{silvio_kac} have estimated a mosaic length scale in
addition to a heterogeneity lenght scale, and have discussed the
crossover in the dominant lengthscale near the mode coupling
temperature. The significantly smaller mosaic length scale they
estimate appears consistent with the smallness of the length scale we
obtain from the finite size dependence of the configurational entropy.
Nevertheless, this analysis does not contain explicit predictions
regarding the relevance of the configurational entropy at temperatures
higher than the mode coupling temperature.  
  
Our observation that the configurational entropy predicts the
relaxation times in accordance with the AG relation for all the
temperatures and system sizes we study poses serious challenges to
current theoretical descriptions based on the analogy with the
behaviour of mean-field models. Although the relevance of the
configurational entropy at high temperatures has been observed in
earlier simulation studies and analyses based on the inherent
structure approach \cite{sastryPRL,sastrynature1,saika}, we emphasize
that a theoretical analysis that satisfactorily explains such
dependence is not at hand at present, and our results concerning
the robustness of the Adam-Gibbs relation in finite systems highlights
further the challenge to existing theoretical descriptions. Equally
importantly, our results reveal that the length scale associated with
dynamical heterogeneity does not play the central role attributed to
it in recent analyses, and highlights the necessity to understand the
role of other relevant length scales, along the lines of the analysis
in \cite{silvio_kac}.


\noindent{\bf Methods :}

\noindent{\it Simulation details:} The system we study is a 80:20 
(A:B) binary mixture of 
particles interacting via the Lennard-Jones potential:
\begin{equation}
V_{\alpha\beta}(r)=4\epsilon_{\alpha\beta}\left[\left(\frac{\sigma_{\alpha\beta}}{r}
\right)^{12}-
\left(\frac{\sigma_{\alpha\beta}}{r}\right)^{6}\right],
\end{equation}
where $\alpha,\beta \in \{A,B\}$ and,
$\epsilon_{AB}/\epsilon_{AA}=1.5$, $\epsilon_{BB}/\epsilon_{AA}=0.5$,
$\sigma_{AB}/\sigma_{AA}=0.80$, $\sigma_{BB}/\sigma_{AA}=0.88$, masses
$m_A = m_B$.  The interaction potential is cutoff at
2.50$\sigma_{\alpha\beta}$. Length, energy and time are reported in
units of $\sigma_{AA}$, $\epsilon_{AA}$ and
$\sqrt{\sigma_{AA}^2/\epsilon_{AA}}$, and other {\it reduced
units} are derived from these. All simulations are done for number
density $\rho = 1.20$. We have used a cubic simulation box with
periodic boundary conditions. Simulations are done in the canonical
ensemble (NVT) using a modified leap-frog integration scheme. We
simulate for seven temperatures in the range $T \in \{0.450,
1.00\}$. The mode coupling temperature for this system has been
estimated \cite{kob} to be $T_{MCT} \simeq 0.435$.  We equilibrate the system for
$\sim 10^7 - 10^8$ MD steps depending on system size and production
runs are at least 5 times longer than the equilibration runs. We use
integration time steps $dt$ from $0.001$ to $0.006$ for the
temperature range $0.800$ to $0.450$.  The studied system sizes vary
from $N = 50$ to $N = 1600$.

\noindent{\it Dynamics:} Dynamics is studied via a two point correlation function, the overlap $Q(t)$, 
\begin{eqnarray}
Q(t)= \int d\vec{r} \rho(\vec{r},t_0)\rho(\vec{r},t+t_0) 
\sim \sum_{i=1}^{N}w(|\vec{r}_i(t_0)-\vec{r}_i(t_0+t)|)
\end{eqnarray} 
where $\rho(\vec{r},t_0)$ {\it etc} are space-time dependent particle
densities, $w(r) = 1$, if $r \le a$ and zero otherwise, and averaging over the
initial time $t_0$ is implied. The use of
the window function [$a = 0.30$] 
treats particle positions separated due to small
amplitude vibrational motion as the same. The second part of the
definition is an approximation which uses only the self-term, which we
have verified to be reliable (see \cite{sharon} for
details). The structural relaxation time $\tau$ is measured by a 
stretched exponential fit of the long-time decay of $Q(t)$.

The fluctuations in $Q(t)$ yields the dynamical susceptibility: 
\begin{equation}
\chi_4(t)=\frac{1}{N}[\langle Q^2(t)\rangle -\langle Q(t)\rangle^2].
\end{equation}
Previous work \cite{sharon} has shown that $\chi_4(t)$ reaches a maximum for times
$\tau_4$ which are proportional to the structural relaxation time $\tau$. We
report the values of $\chi_4^p \equiv \chi_4(t=\tau_4)$. 

The Binder cumulant, which we use for finite size scaling, is defined as
\begin{eqnarray}
B(N,T) &=& \frac{<[Q({\tau_4}) - <Q({\tau_4})>]^4>}{3<[Q({\tau_4}) - 
<Q({\tau_4})>]^2>^2} -1.
\label{BC_def}
\end{eqnarray}
$B(N,T) = 0 $, if the distribution $P(Q({\tau_4}))$ is Gaussian, and is a
scaling function of $\xi/L$ only (where $L$ is the linear size of the
system, and $\xi$ is the correlation length), without any prefactor. 

\noindent{\it Configurational Entropy :} $S_c$, the
configurational entropy per particle, is calculated as the
measure of the number of distinct local energy minima, by subtracting
from the total entropy of the system the `vibrational' component:
\begin{equation}
S_c(T) = S_{total}(T) - S_{vib}(T).
\end{equation} 
Details of the calculation procedure are as given in \cite{sastryPRL}.


\eject 






\centerline{\LARGE \bf Figure Captions}

{\bf Figure 1: System size dependence of dynamic susceptibility 
$\chi_4^p$, and 
finite size scaling of the Binder cumulant $B(N,T)$}.  {\bf a} {\bf Inset:}
$\chi_4(t)$, shown here for $N = 1000$ and selected temperatures,
exhibits non-monotonic time dependence, and the time $\tau_4$ at which it has
the maximum value  has been observed to be proportional to
the structural relaxation time $\tau$. {\bf Main panel:} Peak height of the
four-point dynamic susceptibility, $\chi_4^p(T,N) \equiv
\chi_4(t=\tau_{4}, T, N)$, has been shown as a function of system
size $N$ for different temperatures. For each temperature, $\chi_4^p(T,N)$
increases with system size, and saturates for large system
sizes. $\chi_4^p(T,N)$ also increases as the temperature is lowered. {\bf
b} {\bf Inset:} The distribution $P[Q(\tau_4)-<Q(\tau_4)>]$ 
of $Q(\tau_4)-<Q(\tau_4)>$ is shown for two
system sizes for T = 0.470. While the distribution for the large size
is nearly Gaussian, the small system exhibits a strongly bimodal
distribution. Such bimodality is also observed to emerge as the
temperature is decreased at fixed system size. {\bf Main Panel:}
Binder cumulant B(N,T) (see Methods ) has been plotted as function of
$N/\xi^3$ for different temperatures in the range $T \in [0.45,
0.80]$. The correlation length $\xi$ is an unknown,
temperature dependent, scaling parameter determined by requiring data
collapse for values at different $T$. By construction $B(N,T)$ will
approach zero for large system sizes at high temperatures. It changes
to negative values as the temperature or the system size is decreased
such that $P[Q(\tau_4)-<Q(\tau_4)>]$ becomes bimodal. The correlation length $\xi(T)$ is
the only unknown to be determined in order to obtain data collapse for
$B(N,T)$ and the quality of the data collapse confirms the reliability
of this procedure.

{\bf Figure 2: Relaxation times as a function of temperature and system size}. 
Relaxation time $\tau(T,N)$ for the largest system size increases
roughly by three decades from the highest to the lowest temperature
shown. For each temperature, $\tau(T,N)$ increases as $N$ is decreased
for small values of $N$, 
displaying a trend that is opposite to that observed near
second order critical points. For the smallest temperature,  $\tau(T,N)$ 
increases by about a decade from the largest to the smallest system size.

{\bf Figure 3: Relationship between the relaxation time $\tau(T,N)$,
correlation length $\xi(T)$ and the dynamic susceptibility
$\chi_4^p(T,N)$}. {\bf Inset:} (a) $\tau(T,N \to
\infty)$ is shown against $\xi(T)$, in a log-log plot (bottom curve). This plot shows that a
power-law dependence holds over a temperature range above $T = 0.5$,
but breaks down at lower temperatures. (b) $\tau(T,N \to
\infty)$ is shown against $\xi(T)/k_B T$, in a semi-log plot (top curve). This plot 
shows that an exponential dependence $\tau \sim \exp\left( k (\xi/k_B
T)^\zeta\right)$, with $\zeta = 0.7$, describes the data well in the
entire temperature range. However the observed exponent value $\zeta =
0.7$ is difficult to explain with existing theories. 
{\bf Main Panel:} Relaxation times $\tau(T,N)/\tau(T,N \to
\infty)$ shown against $\chi_4^p(T,N)/\chi_4^p(T,N \to \infty)$ 
in a semi-log plot. While at fixed $N$ both $\tau$ and $\chi_4^p$
increase upon decreasing $T$, at fixed $T$, they show opposite trends,
with $\tau$ increasing for decreasing $N$ and $\chi_4^p$ increasing
for increasing $N$. If $\tau$ and $\chi_4^p$ are determined by the same
length scale $\xi$ and further, if their finite size behavior is
governed by $N/\xi^3$, the plotted data are expected to lie on a
universal curve, which is seen not to be the case.

{\bf Figure 4: The dependence of relaxation times $\tau$ on the
configurational entropy $Sc$}. {\bf Main Panel:} The relaxation times
are shown against configurational entropy in an ``Adam-Gibbs'' plot
[$\log(\tau)$ {\it vs} $1/(TS_c)$], for all the temperatures and
system sizes studied. The impressive data collapse of all the data
onto a master curve indicates that the configurational entropy is
crucial for determining the relaxation times. The overall behavior is
well described by the AG relation, which requires $\log(\tau)\propto
1/(TS_c)$. Although other powers of $1/(TS_c)$ may also describe the
data well, the improvement in fit quality is marginal, and hence we
treat the data presented as validating the AG relation. {\bf Top
Inset:} The configurational entropy $S_c(T,N)$ scaled to its $N
\rightarrow \infty$ value, has been plotted as function of $N/\xi_s^3$
for different temperatures in the range $T \in [0.45, 0.80]$, to
extract a temperature dependent length scale $\xi_s$ that leads to
data collapse. {\bf Bottom Inset:} The length scale obtained from the
data collapse of the configurational entropy (green diamonds) compared
with the length scale obtained from finite size scaling of the Binder
cumulant (red triangles). It is apparent that the length scale from
configurational entropy shows very weak temperature dependence, in
contrast with the dynamical heterogeneity length scale.

\eject


\begin{figure}[h]
\begin{center}
\epsfig{file=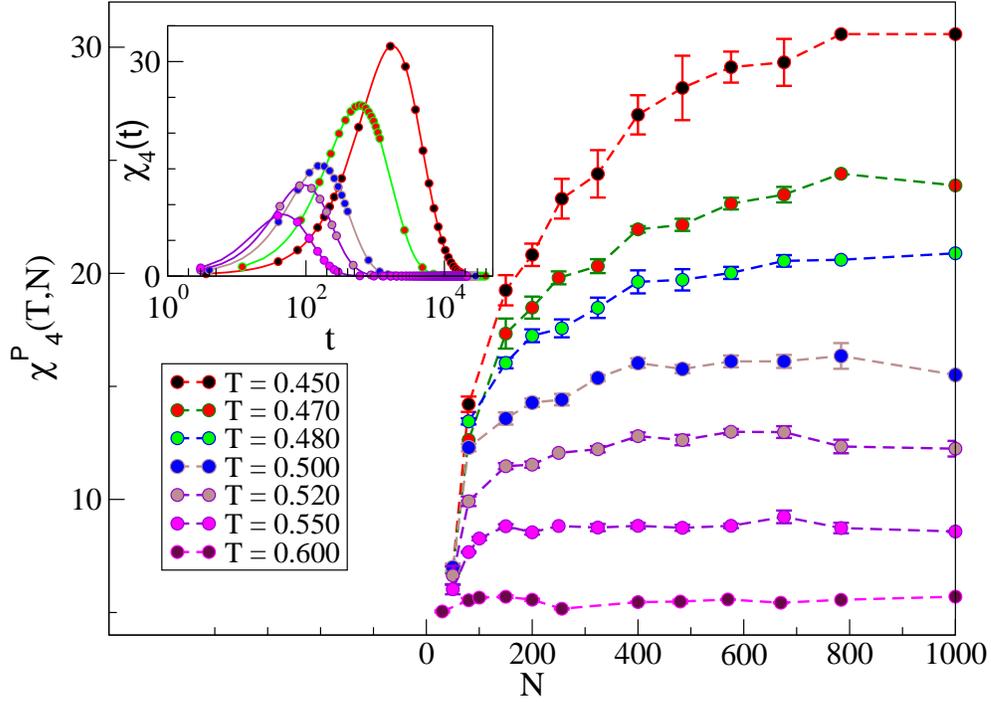,scale=0.5,angle=-90,clip=} \epsfig{file=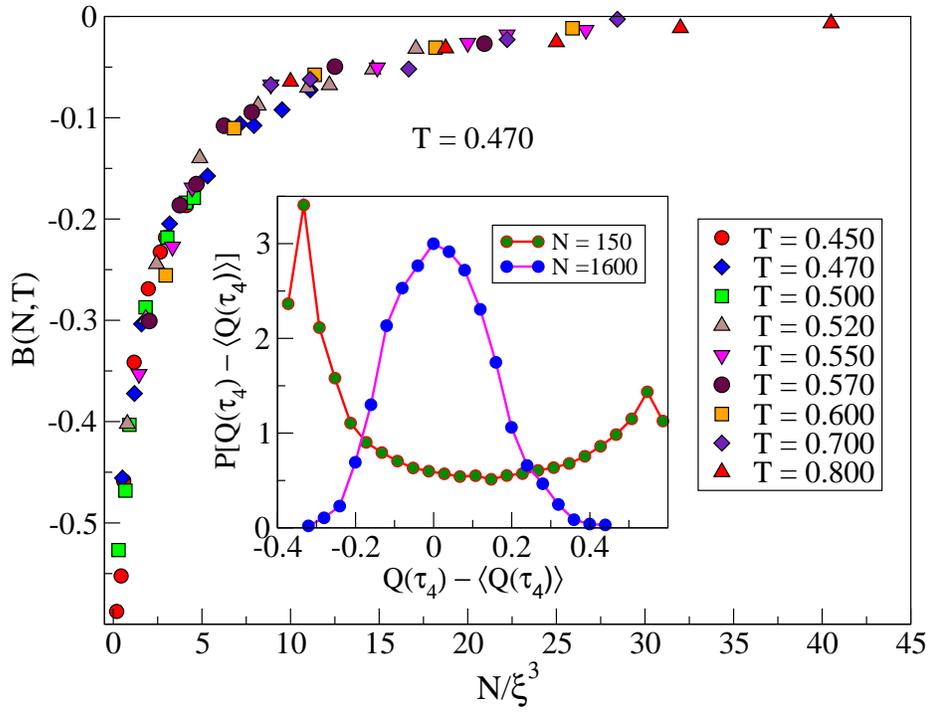,scale=0.5,angle=-90,clip=}
\caption{\bf Karmakar et al}
\end{center}
\end{figure}

\pagebreak 

\begin{figure}[h]
\begin{center}
\epsfig{file=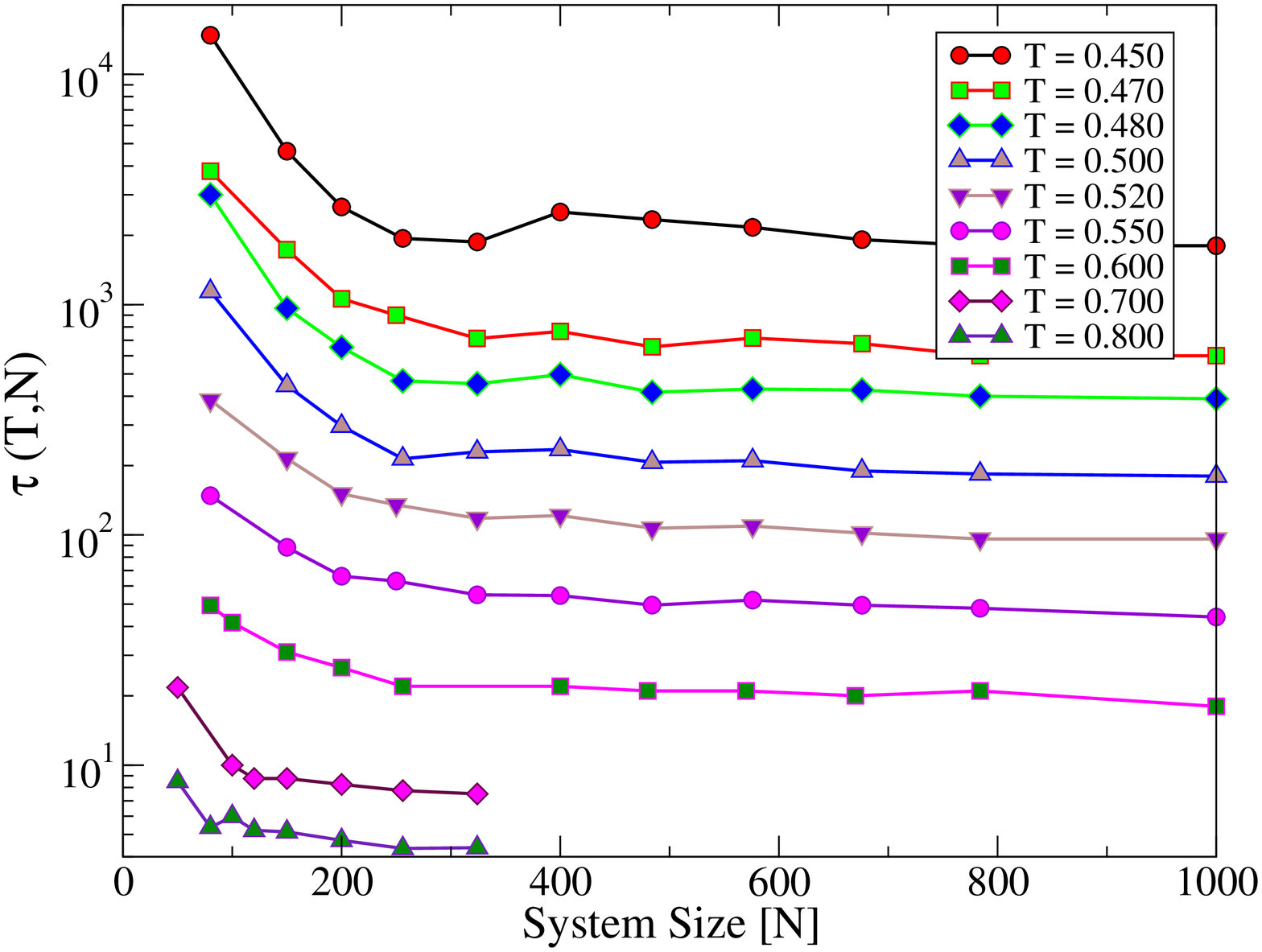,scale=0.5,angle=-90,clip=}
\caption{\bf Karmakar et al}
\end{center}
\end{figure}

\pagebreak 

\begin{figure}[h]
\begin{center}
\epsfig{file=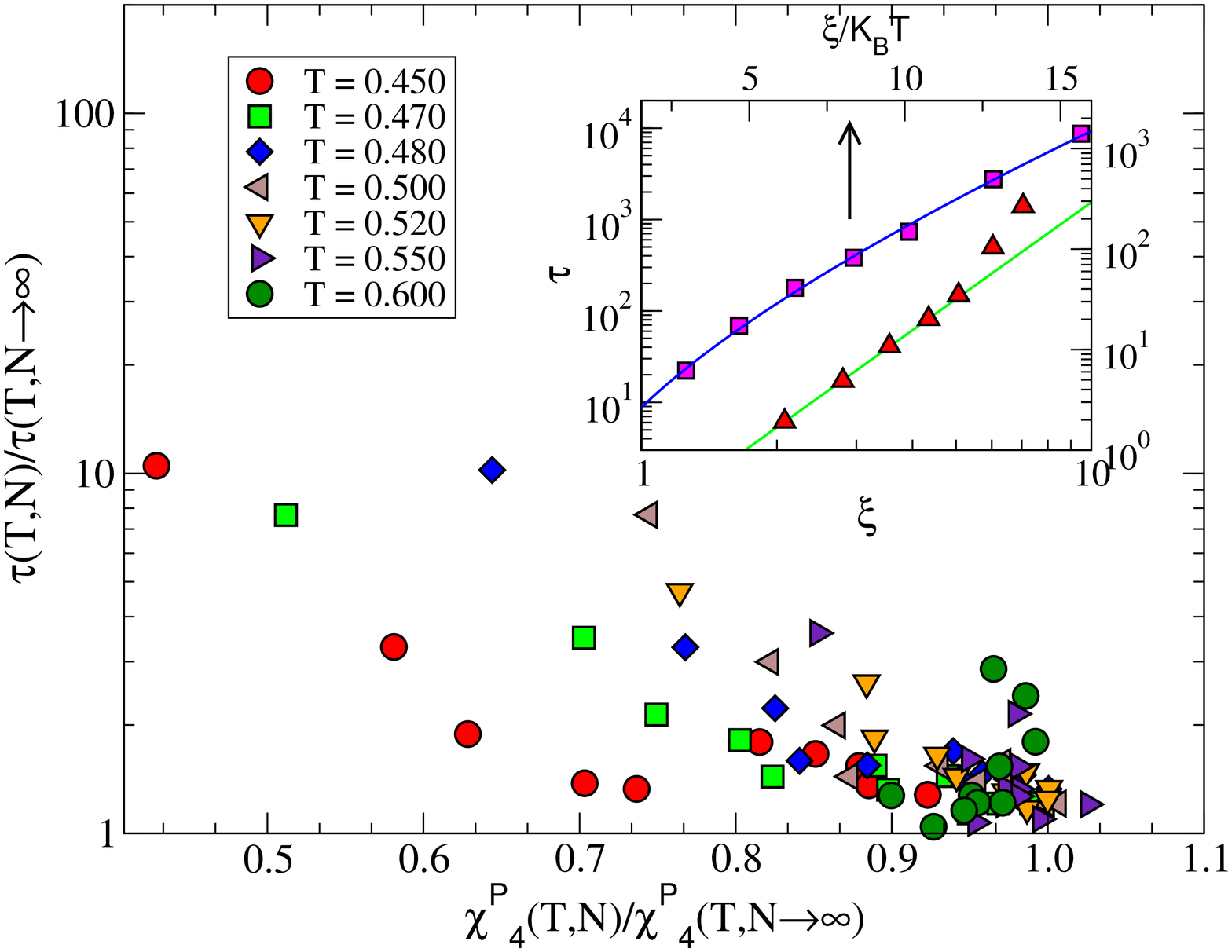,scale=0.5,angle=-90,clip=}
\caption{\bf Karmakar et al}
\end{center}
\end{figure}

\pagebreak

\begin{figure}[h]
\begin{center}
\epsfig{file=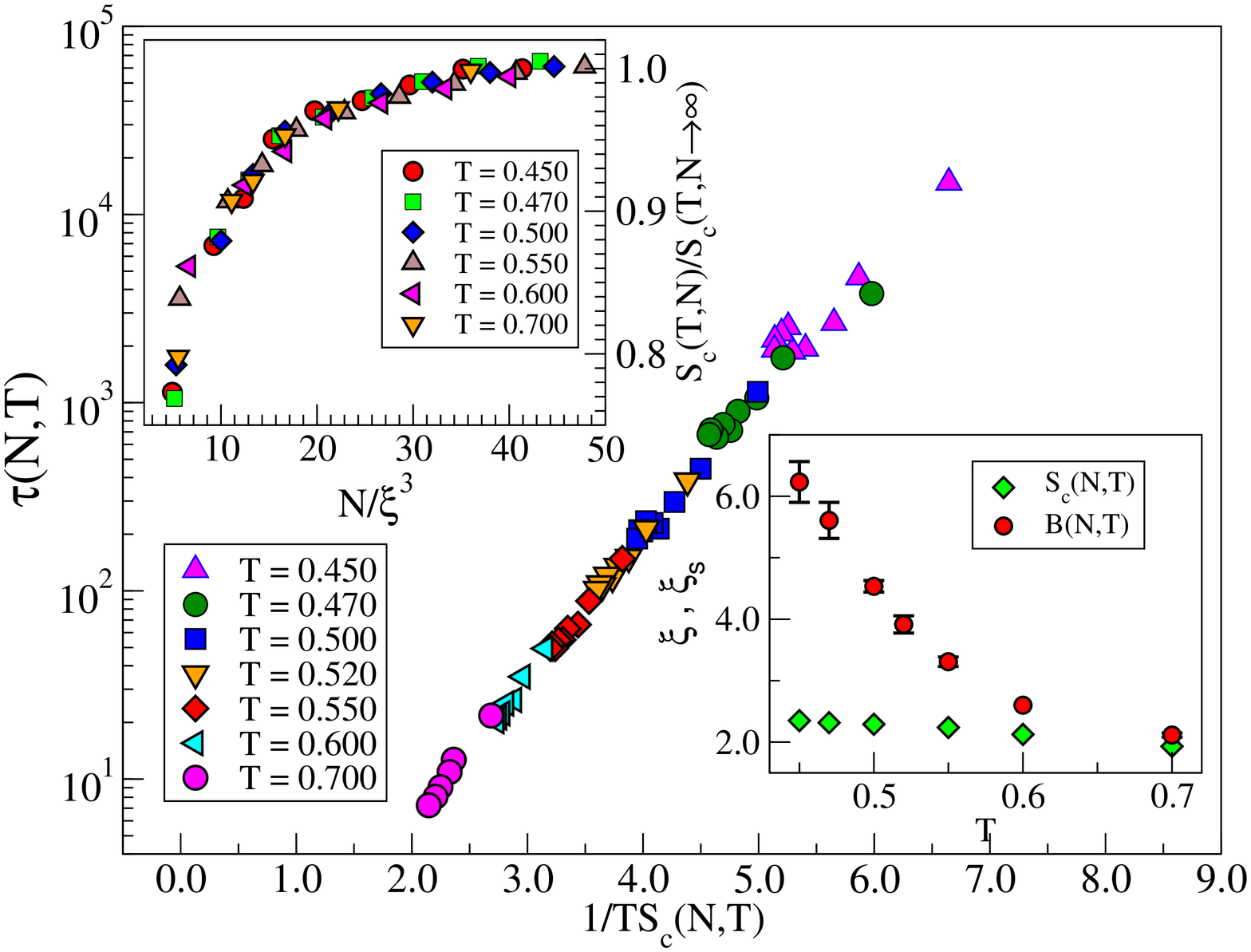,scale=0.5,angle=-90,clip=}
\caption{\bf Karmakar et al}
\end{center}
\end{figure}

\end{document}